\documentclass[twocolumn,prb,showpacs]{revtex4}

\usepackage{graphicx}
\usepackage{rotating}
\usepackage{amsmath}
\usepackage{amsfonts}
\usepackage{amssymb}
\usepackage{enumerate}
\usepackage{longtable}
\usepackage{dcolumn}
\usepackage{bm}

\begin{document}

\newcommand{\be}{\begin{equation}}
\newcommand{\ee}{\end{equation}}
\newcommand{\bn}{\begin{eqnarray}}
\newcommand{\en}{\end{eqnarray}}
\newcommand{\hg}{${\rm Hg_{2}Ru_{2}O_{7}}$}

\title{Self-Doping Induced Orbital-Selective Mott Transition in 
\hg}

\author{L. Craco,$^{1,2}$ M.S. Laad,$^2$ S. Leoni,$^1$ and H. Rosner$^1$}
\affiliation{
$^1$Max-Planck-Institut f\"ur Chemische Physik fester Stoffe, 
01187 Dresden, Germany \\
$^2$Max-Planck-Institut f\"ur Physik komplexer Systeme, 
01187 Dresden, Germany} 

\date{\rm\today}

\begin{abstract}
Pyrochlore oxides are fascinating systems where strong, multi-orbital 
correlations in concert with geometrical frustration give rise to 
unanticipated physical properties. The detailed mechanism of the 
insulator-metal transitions (IMT) underpinning these phenomena is, 
however, ill-understood in general. Motivated thereby, we study the 
IMT in the pyrochlore ${\rm Hg_{2}Ru_{2}O_{7}}$ using  LDA+DMFT. In 
contrast to the well-known examples of Mott transitions in TMO, we 
show that, in the negative charge-transfer situation characteristic 
of \hg, self-doping plays a crucial role in the emergence of an 
orbital-selective IMT. We argue that this mechanism has broader 
relevance to other correlated pyrochlore oxides.   
\end{abstract}
     

\maketitle

\section{INTRODUCTION}

The Mott-Hubbard insulator-metal transition (IMT) is by now recognized 
to play a central role in our understanding of $d$- and $f$-band 
compounds.~\cite{[1]} Understanding the complex interplay between 
strong, multi-orbital (MO) electronic correlations, structural distortions, 
and strongly anisotropic, orbital-dependent hopping holds the key to a 
consistent theoretical description of their unique responses. Adding 
geometric frustration to the above results in a truly formidable 
problem. In geometrically frustrated systems, the exponentially large 
degeneracy of classical ordered states inhibits emergence of conventional 
order, permitting new, complex ordered ground states to arise.~\cite{[2]} 
In {\it real} systems, structural effects may partially remove this huge 
degeneracy, making the problem (counter-intuitively) somewhat simpler 
to solve;~\cite{[3]} however, in near-undistorted cases, near perfect 
orbital degeneracy, and the consequent strong quantum orbital {\it and} 
spin fluctuations in a highly degenerate system underpin their physical 
behavior.

The recently discovered pyrochlore system, \hg,~\cite{[4]} is a 
particularly interesting case in this context. As temperature $(T)$ is 
reduced, the anomalous (see below) non-Fermi liquid (nFL) metallic state 
becomes unstable, via a first-order Mott transition, to an antiferromagnetic 
Mott-Hubbard insulator (AFI).~\cite{yamamoto} External pressure ($p$) melts 
the AFI beyond a critical $p_{c}=6.0$~GPa, resulting in a low-$T$ correlated 
FL behavior for $T<T^{*}=13$K. At ambient pressure, the IMT transition 
is accompanied by lowering of lattice symmetry from (high-$T$) cubic to 
(low-$T$) a lower, hitherto precisely unknown, type: depending upon its 
precise type, either AF or dimer order may be possible.~\cite{[5]}  


The high-$T>T^{*}$ state in \hg~is an anomalous nFL: the $dc$ 
resistivity, $\rho(T) \simeq T$ for $T>T_{MI}=108$~K~\cite{[4]} at 
ambient pressure, and deviates from the FL form for $T>T^{*}$ beyond $p_{c}$. 
The uniform spin susceptibility is Curie-Weiss like for $T>T_{MI}$, 
indicating a strong local moment scattering regime. The nFL character 
is also borne out from the recent photoemission (PES) data,~\cite{[6]} 
showing anomalously broad PES lineshapes, with no hint of FL quasiparticles, 
in the metallic phase. Given the cubic pyrochlore structure for $T>T_{MI}$, 
strong orbital (from $t_{2g}$ orbital degeneracy) and spin fluctuations are 
implied. How might strong scattering processes involving these fluctuations 
produce the observed nFL metal? What drives the AFI as $T$ is lowered? A 
correlated electronic structure study which can illuminate these issues 
does not, to our best knowledge, exist. In this work we study precisely 
these issues in part in \hg~using the LDA+DMFT method.~\cite{held}  We focus 
on the mechanism of the $T$-driven IM transition, and leave the issue of the 
low-dimensional AF with a spin gap for future consideration.

\section{MODEL AND SOLUTION}

\begin{figure}[thb]
\includegraphics[width=3.3in]{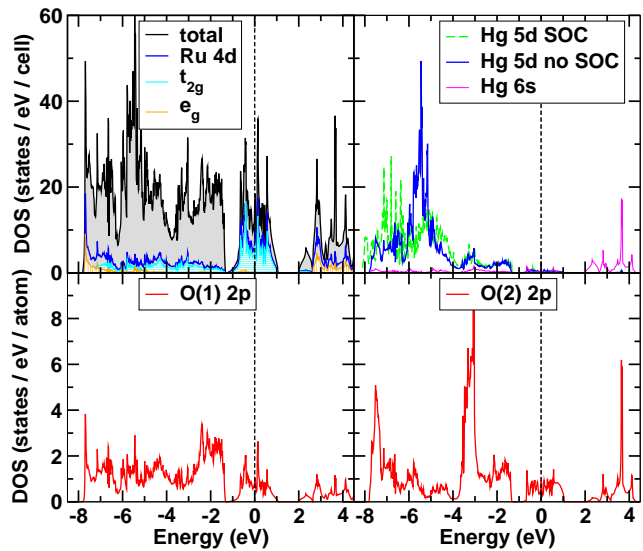}
\caption{(Color online) 
LDA band structure for cubic \hg. $O(1)~(O(2))$ denotes Oxygen ions 
nearest to Ru (Hg).}
\label{fig1}
\end{figure}

Starting with the high-$T$ cubic $Fd\bar 3m$ structure 
found by Klein {\it et al.}~\cite{[4]} local density approximation (LDA) 
band structure calculations were performed using a (scalar and 
fully-relativistic) full-potential local-orbital scheme (FPLO)~\cite{fplo} 
and a linear muffin-tin orbitals (LMTO) scheme in the atomic sphere 
approximation.~\cite{ok} In Fig.~\ref{fig1} we display our 
FPLO-LDA~\cite{fpcoment} results for the one-particle 
density-of-states (DOS). Clearly, the major contribution to the DOS 
comes from $Ru(4d), O(1)(2p)$ orbitals, but the $O(2)(2p)$ orbitals 
also have noticeable spectral weight at the Fermi level $(E_{F})$. As 
seen in Fig.~\ref{fig1}, the influence of the spin-orbit coupling (SOC) 
in the cubic phase to the $t_{2g}$ states near $E_F$ is negligible. 
Further, we observe strong hybridization between $Ru(4d)-O(1)(2p)$ 
orbitals, and between $O(2)(2p)-Hg(6s)$ orbitals, but weak mixing 
between these two sets.  These important findings will be exploited below to 
study the physics of \hg using a multi-orbital Hubbard model involving only
the $d$-band sector.  At one-particle level, the corresponding model 
Hamiltonian reads,
$H_{band}=\sum_{k,a,\sigma}\epsilon_{ka}d_{ka\sigma}^{\dag}d_{ka\sigma} 
+ \sum_{i,a,b=1,2} \Delta_{b}(n_{i,b}^p-n_{i,a}^d),$
where $a$ labels the diagonalized combination of 
($e_{g1}',e_{g2}',a_{1g}$)~\cite{diagonal} 
the $t_{2g}$ orbitals, $\Delta_{b}~(b=1,2)$ are the $pd$ charge transfer 
(CT) terms involving two $Ru-O(1),(b=1)$ and $Ru-O(2),(b=2)$ channels. 
Clearly, neither an AFI Mott-Hubbard insulator nor a nFL metal can be 
expected at this level, this requiring a reliable treatment of strong, 
$d$-shell electronic correlations. This part reads,

\be
\nonumber
H_{int}=U\sum_{i,a}n_{ia\uparrow}^d n_{ia\downarrow}^d 
+ U'\sum_{i,a,a'}n_{ia}^d n_{ia'}^d 
- J_{H}\sum_{i,a,a'}{\bf S}_{ia} \cdot{\bf S}_{ia'}
\ee
with $a,a'=e_{g1}',e_{g2}',a_{1g}$. Given the larger spatial 
extent of $4d$-orbital vis-a-vis their $3d$ counterpart, we also include a 
Madelung term, $H_{M}=U_{pd}\sum_{<i,j>,a,b}n_{i,a}^d n_{j,b}^p$ in our 
calculations (see below). 
In ${\rm Tl_{2}Mn_{2}O_{7}}$, the IMT (from a paramagnetic insulator (PI) 
to a ferromagnetic metal (FMM)) is seemingly driven 
by the $Tl(6s)$ states crossing $E_{F}$ across $T_{c}$,~\cite{[9]} for 
instance. This can occur via a $T$-dependent CT from the TM $d$-states 
to the $O(2)-Tl$ hybridized states. In view of the generic relevance of 
{\it self-doping} in TM $4d$ pyrochlores,~\cite{[8],[9]} this term is an 
essential part. In the low-T phase, this will involve CT processes involving 
the $O(2p)-Hg(6s)$ channel (second term of $H_{band}$) across the IMT, as we 
describe below.

The $d_{a\sigma}$ above should be understood as appropriate, $RuO_{6}$ 
cell-centered, combinations of the Ru-$4d$ and O-$2p$ orbitals, computed 
within LDA.  Similarly, the $b=2$ channel is to be understood as a band of 
hybridized O-$2p$ and Hg-$6s$ orbital states.  Because of the negligible 
one-particle mixing (hybridization) between the $b=1,2$ channels, we 
approximate the full problem of three Ru-$d$ bands coupled to the $b=2$ band
channel by replacing the latter by a ``reservoir'', whose only
function is to simulate the self-doping process arising from the negative charge
transfer situation that obtains in \hg.  Of course, this is an approximation.
It is, however, a good one: the $b=2$ band channel has appreciably smaller DOS 
around $E_F$ in the LDA results.  In a DMFT-like approximation, with negligible
one-particle hybridization between the $b=1,2$ bands, the {\it intersite}
Madelung term will push this small spectral weight away from $E_F$, to 
lower- and higher energy (notice that the O-$2p$ bands will be split by 
$U_{pd}z\langle n_{d}\rangle$, where $z$ is the co-ordination number of 
the lattice, and hence quite large). This will already occur at the level 
of LDA+Hartree approximation. We then expect that the {\it correlated} 
spectral function will be dominated by the $d$-bands over an appreciable 
range about $E_F$.  Correlation effects in \hg~via LDA+DMFT are studied 
below, subject to this caveat; we will show that this is indeed a good 
approximation {\it a posteriori}, in the sense that our LDA+DMFT results 
with the above caveat show very good quantitative agreement with 
one-particle spectroscopy and key thermodynamic and transport data in \hg.  

The full many-body Hamiltonian reads $H=H_{band}+H_{int}+H_{M}$. We solve 
this model within MO dynamical mean field theory (MO-DMFT) developed and 
used for a range of TMO with good success.~\cite{held} We use the 
MO-iterated perturbation theory (IPT) as an impurity solver in the DMFT 
selfconsistent procedure. Though not numerically exact (like QMC, NRG, 
D-DMRG), it has many advantages: it is valid for $T=0$, where QMC cannot 
be used. NRG/DDMRG are extremely prohibitive for three orbital models, 
even without electronic structural inputs at LDA level.  As shown in 
earlier work,~\cite{[11]} DMFT(MO-IPT) generically gives very good 
semiquantitative agreement with PES/X-ray absorption (XAS) experiments for 
TMOs.~\cite{[11],[miura]} For \hg, 
DMFT(MO-IPT) has to be extended to treat the second CT channel described 
above. Given the complexity of the problem, we choose the following 
strategy to solve $H$ above: For the high-$T$ phase, we $(i)$ solve 
$H_{band}+H_{int}$ within DMFT(MO-IPT) version used earlier, for technical 
details see Ref.~\onlinecite{[11]}. 
$(ii)$ As indicated by experiment on the related 
${\rm Tl_{2}Ru_{2}O_{7}}$ system,~\cite{[8]} we study the effect of 
$Ru(4d)-O(2)(2p)-Hg(6s)$ CT processes by incorporating this CT channel 
selfconsistently into the LDA+DMFT(MO-IPT) procedure: given the small 
$Ru(4d)-O(2)(2p)$ {\it hybridization}, the $O(2)(2p)-Hg(6s)$ 
channel acts like a {\it scattering} (non-hybridizing) channel for the 
$Ru-t_{2g}$ bands in the impurity model of MO-DMFT. This enables us to 
treat this extra channel by combining the MO-IPT solution for $(i)$ with 
the {\it exact} DMFT solution of a Falicov-Kimball model (FKM) for 
$(ii)$,~\cite{[12]} in a selfconsistent way.

\section{LDA+DMFT RESULTS AND DISCUSSION}

We now describe our LDA+DMFT results. We start with the (cubic) pyrochlore 
structure at high-$T$, with the corresponding LDA DOS as the input for 
the MO-DMFT calculation with total $4d$ occupation, $n_{t}=3$. Further, 
we work in the (LMTO)~\cite{sphere} representation in 
which the one-particle density matrix is diagonal in the $t_{2g}$ orbital 
index, so that $G_{\alpha\beta}^{(0)}(k,\omega)=\delta_{\alpha\beta}
G_{\alpha\alpha}^{(0)}(k,\omega)$. 
We choose $U=5.5$~eV, $J_{H}=1.0$~eV, and $U'\simeq (U-2J_{H})=3.5$~eV 
for the $Ru-4d$ shell, along with $U_{pd}\simeq 1.5$~eV, as representative 
values for \hg. We believe that $U,~J_H$ do not vary much for $4d$-TMO 
pyrochlores, and, in fact, $U=5.0$~eV was found for the CT insulator 
${\rm Cs_2AgF_4}$.~\cite{helge}

\begin{figure}[thb]
\includegraphics[width=3.5in]{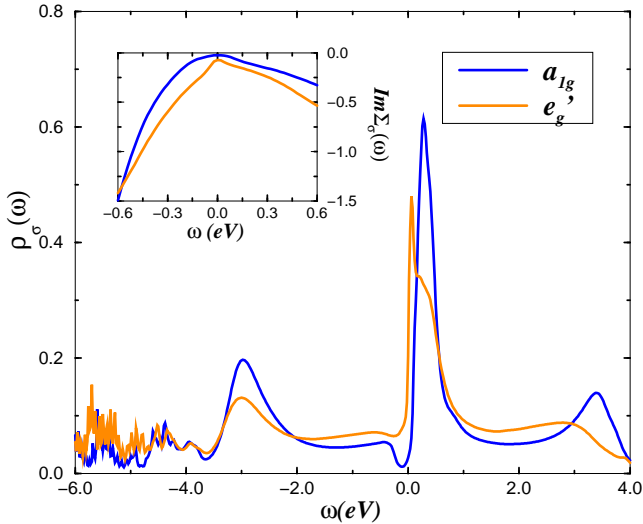}
\caption{(Color online) 
$t_{2g}$-resolved LDA+DMFT densities of states for \hg in the high-$T$ 
nFL phase, for $U=5.5$~eV and $J_H=1$~eV. Notice the orbital selective 
nature of the metal, as well as the nFL character in the orbital resolved 
self-energies (see inset).}
\label{fig2}
\end{figure}

\subsection{Metallic Phase}

In Fig.~\ref{fig2}, we show the correlated many-particle spectral function 
for the metallic phase of \hg.  The dynamical spectral weight transfer (SWT) 
over large energy scales, characteristic of strong, local correlations is 
explicitly manifest.  More interestingly, the metal has an 
{\it orbital selective} (OS) character: the $a_{1g}$ DOS is almost ``Mott'' 
localized, even as the $e_{g}'$ DOS develops a precursor of a low-energy
pseudogap, characteristic of an incoherent metal behavior. This is further 
corroborated by examining the orbital dependent self-energies 
($\Sigma (\omega)$), see inset in Fig.~\ref{fig2}. Clearly, the $e_{g}'$ 
(imaginary part of) $\Sigma (\omega)$ shows {\it quasi-linear} frequency 
dependence near the Fermi energy ($|\omega-E_{F}|<$0.3~eV), while 
Im$\Sigma_{a_{1g}}(\omega)$ shows quadratic behavior. This describes a 
nFL metal, with a linear-in-$T$ quasiparticle damping rate. 
Within DMFT, this is also the transport relaxation rate,~\cite{[georges]} 
since vertex corrections drop out in the computation of the conductivities 
in this limit. The $dc$ resistivity is then given by 
$\rho_{dc}(T)\simeq (m^{*}/ne^{2})$Im$\Sigma_{e_{g}'}(\omega=T)\simeq AT$, 
in accord with the linear-in-$T$ resistivity observed experimentally in 
the ``high-$T$'' metallic phase. Moreover, the selective localization 
seen in the $a_{1g}$ orbital DOS gives rise to unquenched local momemts 
co-existing with ``itinerant'' (but incoherent as derived above) $e_{g}'$ 
carriers naturally gives a Curie-Weiss form of the spin susceptibility, 
which is also observed right up to the IMT. We have also estimated the 
$\gamma$-co-efficient of the metallic specific heat from the real part 
of $\Sigma_{\alpha}(\omega)$ (not shown) as $\gamma/\gamma_{LDA}=4.25$: 
this seems to be in the range estimated in Ref~\onlinecite{[4]}. Actually, 
the noticeable $T$ dependence of $\gamma$ above $T_{MI}$~\cite{[4]} is 
additional evidence of disordered local moments in the ``bad'' metal, 
and further supports our picture.

\begin{figure}[thb]
\includegraphics[width=3.5in]{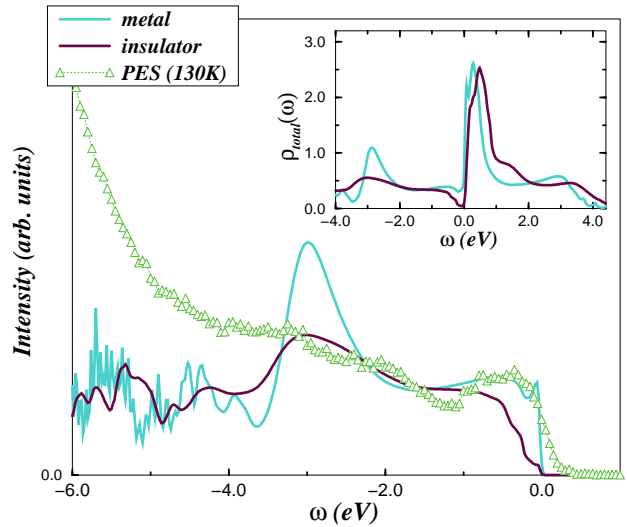}
\caption{(Color online) 
Theoretical PES spectra (LDA+DMFT DOS convoluted with instrumental 
resolution) in the nFL metallic and insulating phases of \hg. In the 
LDA+DMFT calculation, the total $t_{2g}$ electron density changes from 
$<n_t>=3.0$ (metallic) to $<n_t>=2.6$ (insulating). The theoretical 
result shows very good agreement with experimental PES result in the 
metallic phase, taken from Ref.~\onlinecite{[6]} up to 2~eV binding energy. 
The inset shows the total spectral function.}
\label{fig3}
\end{figure}
  
Using the LDA+DMFT result, we also compare (Fig.~\ref{fig3}) our computed PES 
lineshape, with very recent work from the RIKEN-SPring 8 group.~\cite{[6]} 
Given our restriction to the $t_{2g}-Ru(4d)$ bands in the MO-DMFT 
($Hg(d), O(2p)$ bands will start contributing at higher binding energies, 
as seen from LDA), good quantitative agreement with experiment is evident 
up to $-2.0$~eV, lending strong support to our theoretical work. Additionally, 
we predict that an intense peak will be seen around $0.4-0.5$~eV in X-ray 
absorption (XAS) studies, as in ${\rm Tl_{2}Ru_{2}O_{7}}$.~\cite{[17]}

In fact, the LDA+DMFT spectral functions show that, at low energy the full 
MO problem is mapped onto an effective FKM-like model~\cite{[11]}, since the 
bad-metal is of the orbital-selective type. There, 
the problem in the local limit corresponds to itinerant (but incoherent) 
$e_{g}'$ carriers scattering off (Mott) localized $a_{1g}$ electronic states. 
The resulting problem is precisely the ``X-ray edge'' problem in 
DMFT.~\cite{[14]} We now understand the structure of the self-energies, 
and the nFL behavior, as an interesting manifestation of the Anderson OC 
caused by this X-ray-edge mapping. Building upon this understanding, using 
DMFT, we predict that, for $T>T_{MI}$: $(a)$ the optical conductivity will 
show a low-energy pseudogapped form, characteristic of an incoherent (nFL) 
metal. Polarised optical studies should indicate the OS-character of the 
nFL state, and, $(b)$ given the local version of the Shastry-Shraiman 
relation,~\cite{[13]} the electronic Raman lineshape should show a 
continuum response, cut off by a pseudogap feature at low energy.

\subsection{Insulating Phase}

Now, we turn to a description of the insulating phase. A naive search for 
the instability of the nFL to a Mott-Hubbard insulating state, where $U,U'$ 
were increased to unphysical values, nevertheless turned out to be 
unsuccessful. This is a clear indication of involvement of additional, 
electronic-cum-structural effects in driving the IMT. The observed structural 
change across $T_{MI}$~\cite{[4],yamamoto} supports this reasoning. 
From the LDA orbital assignment, it is clear that an additional structural 
change necessarily involves {\it partial} occupation of the two-fold 
degenerate $e_{g}'$ orbitals. Starting from the nFL-metal derived above, 
this can only occur via a $T$-dependent change in the ``self-doping'' 
process.~\cite{[8]} As in ${\rm Tl_{2}Ru_{2}O_{7}}$, this $T$-dependence 
could be provided by electronic coupling to the $Ru-O$ stretching phonon 
mode, which is experimentally observed to split below 
$T_{MI}$.~\cite{[8]} This would imply a structural change setting in 
below $T_{MI}$, whose precise nature is hitherto unknown.

Motivated by this observation, we argue that this resulting change in the
$Ru(4d)\rightarrow O(2)(2p)$ CT leads to a partial $4d$ occupation, 
induces orbital order, and lifts the $e_{g}'$ degeneracy via a structural 
distortion. Within DMFT, this will reduce the orbital-dependent hoppings, 
driving large SWT from low- to high energy, and stabilize the second, 
Mott-Hubbard insulating, solution of the DMFT equations. This is indeed 
seen in the LDA+DMFT calculation, as we show below.

Based on this reasoning, we explore the Mott-Hubbard insulating phase of \hg by
searching numerically for the instability of the first (metallic) solution
found above to the second, insulating solution of the DMFT equations
in the quantum paramagnetic phase. The 
DMFT(MO-IPT) equations are solved with $U,U',U_{pd}$, and $n_{t}=2.6$.
We vary $n_{t}$ in trial steps, and look for a critical $n_{t}^{(c)}$ which
stabilizes the PI solution of the DMFT equations.  
As discussed above, and indeed seen in the inset of Fig.~\ref{fig3}, partial 
occupation of the 
$4d-{e_{g}'}$ orbitals leads to removal of the orbital degeneracy, and the
corresponding modification of orbital orientation gives rise to a reduced 
{\it intersite} one-electron 
overlap.  Within MO-DMFT, this triggers the Mott-Hubbard insulating state via 
large SWT on a scale of $5.0$~eV, as seen in Fig.~\ref{fig2}. Hence, the IMT is 
an OS Mott transition.  Importantly, however, notice that the self-doping process 
leading to fractional $d$-orbital occupation is a crucial ingredient.  Thus,
in contrast to other OS cases,~\cite{[11]} the IMT 
in \hg~is driven by the self-doping in the negative CT situation, 
$\Delta_{2}=(\epsilon_{p2}-\epsilon_{d})$=(-2.8eV+1.3eV)=-1.5eV. Along 
with the microscopic elucidation of the nFL behavior, the  
agreement with PES in the nFL metal phase up to $\omega \simeq -2.0$~eV
(Fig.~\ref{fig3}) constitutes strong evidence in favor of our mechanism 
for the IMT in \hg. Quantitative comparison with PES in the low-$T$  
phase requires an extension of our approach to include short-ranged 
(intersite) orbital and spin correlations characteristic of pyrochlores. 
This requires a cluster-DMFT analysis, presently a forbidding prospect.
Nevertheless, observation of a Curie-Weiss spin susceptibility right up 
to the IMT,~\cite{[5],[variousp]} along with the excellent agreement with 
PES in the nFL, justifies the use of DMFT to describe the IMT.

We emphasize that this is a new picture for the IMT in correlated 
systems. In contrast to the early TMO, which are Mott-Hubbard insulators 
($U_{dd} > \Delta$ in the Zaanen-Sawatzky-Allen scheme),~\cite{[15]} the 
$Ru(4d)-O(1)(2p)$ CT channel, along with the Madelung term, are 
important in \hg. In the DMFT context, the importance of the CT energy 
in the late-TMO is recognized.~\cite{[16]} Our work is the first of its 
kind showing its relevance for the IMT in a MO pyrochlore system. Further, 
it is likely to be more broadly generic to pyrochlore TMOs; recall that, in 
${\rm Tl_{2}Mn_{2}O_{7}}$, the PI-FMM~\cite{[9]} transition is driven by the 
shift of the $Tl(6s)$ band through $E_{F}$ across $T_{c}$. This is 
readily rationalized in our picture, in terms of the ``switching on'' 
the $Mn(3d)-O(2)(2p)-Tl(6s)$ CT channel across $T_{c}$ in 
${\rm Tl_{2}Mn_{2}O_{7}}$.~\cite{[9]}

The correlation between the CT process detailed above and the IMT is also 
visible in a whole family of $4d$ pyrochlores, 
${\rm A_{2}Ru_{2}O_{7}}$, with $A=Pb,~Bi,~Y$.~\cite{[17]} The low-$T$ 
magnetic ordering in the Mott-Hubbard insulating phase(s) may, however, 
be quite different, being sensitively dependent on the nature of the 
structural change across the IMT, as well as on spin state. For example, 
the spin $S=1$ system, ${\rm Tl_{2}Ru_{2}O_{7}}$, shows a very similar 
Mott transition; however, the low-$T$ phase is found to be a Haldane spin 
chain.~\cite{[18]} In \hg, the half-integer spin $S=3/2$ rules out the 
Haldane analogy. If the low-$T$ magnetic structure corresponds to having 
$Ru$ chains, as in ${\rm Tl_{2}Ru_{2}O_{7}}$, one would 
have an AF ground state with gapless spin excitations. Observation of 
the spin-gap in the uniform spin susceptibility in \hg~may therefore point 
to the relevance of the spin-orbit coupling in the {\it insulating} phase: 
this will induce Ising-like anisotropy in an $S=3/2$ Heisenberg chain and 
generate a gap to spin excitations.~\cite{[19]}  More experimental work 
is also called for to pinpoint the specific factors affecting this issue. 
Given the structural distortion necessarily accompanying the IMT caused by 
lifting of the $e_{g}'$ degeneracy in \hg, more detailed theoretical work 
awaits more precise characterization of the low-$T$ structure of \hg. 
We plan to address the issue of magnetic order and its associated excitation 
spectrum in the low-$T$ (insulating) phase of \hg~in a future work.
 
\section{CONCLUSION}

In conclusion, we have performed LDA+DMFT calculations to demonstrate 
the role of multi-orbital electron-electron interactions on the electronic 
structure of a recently discovered pyrochlore system, \hg.  This system
exhibits a first-order temperature dependent IMT. The high-T metallic 
phase is shown to be an orbital selective non-Fermi liquid, signalled by 
a quasi-linear frequency dependence of the imaginary part of 
the correlated self-energy and the absence of quasi-particle peaks in 
the orbital-selective spectral functions at very low frequencies. The 
LDA+DMFT spectral function shows good quantitative agreement 
with recent photoemission (PES) data, showing anomalously broad PES 
lineshapes, as well as with the linear-in-$T$ resistivity and mass 
enhancement in the metallic phase. In stark contrast to the better 
known examples~\cite{[1]} of Mott transitions in $3d$-transition 
metal compounds, our results imply a new mechanism for MIT in the 
$4d$ pyrochlore-based electron systems (like \hg). Namely, a negative 
charge-transfer associated self-doping drives an orbital-selective 
MIT at low temperatures via the Mott-Hubbard route. Our study should 
be more generally applicable to MO pyrochlore systems, and, in
particular, to ${\rm Tl_{2}Mn_{2}O_{7}}$, exhibiting Mott-Hubbard 
transitions~\cite{[8],[17],[variousp],more} as functions of 
suitable ``tuning parameters''.

\acknowledgements  
We are indebted to A. Chainani for discussions, and, especially, for 
sending us his unpublished PES results. LC and HR thank the Emmy-Noether 
Program of the DFG for support.


\begin{thebibliography}{28}
\bibitem{fpcoment} Self-consistency was obtained on a 12x12x12 $k$ mesh 
for the full Brillouin zone.
\end{thebibliography}
\end{document}